\documentclass[aps,prl,twocolumn,superscriptaddress,showpacs,preprintnumbers,amsmath,amssymb]{revtex4}

\usepackage{graphicx} 
\usepackage{dcolumn}  
\usepackage{epstopdf}
\usepackage{amsmath}
\usepackage{units}
\usepackage{verbatim}
\usepackage{xcolor}
\RequirePackage{xspace}
\usepackage{relsize}
\def\babar{\mbox{\slshape B\kern-0.1em{\smaller A}\kern-0.1em B\kern-0.1em{\smaller A\kern-0.2em R}}}
\def\ccbar {\ensuremath{c\overline c}\xspace}
\def\Kbar  {\kern 0.2em\overline{\kern -0.2em K}{}\xspace}

\def\Kzb   {\ensuremath{\Kbar^0}\xspace}
\def\KS    {\ensuremath{K^0_{\scriptscriptstyle S}}\xspace} 
 
\def\Dbar  {\kern 0.2em\overline{\kern -0.2em D}{}\xspace}

\def\Dzb   {\ensuremath{\Dbar^0}\xspace}
\def\Dstar {\ensuremath{D^*}\xspace}

\def\Dstarp{\ensuremath{D^{*+}}\xspace}
\def\Dstarm{\ensuremath{D^{*-}}\xspace}
\mathchardef\Upsilon="7107
\def\Y#1S{\ensuremath{\Upsilon{(#1S)}}\xspace}
\def\cm    {\ensuremath{{\rm \,cm}}\xspace}
\def\invfb {\ensuremath{\mbox{\,fb}^{-1}}\xspace}
\def\order {{\ensuremath{\cal O}}\xspace}
\def\CP    {\ensuremath{C\!P}\xspace}
\def\FB    {\ensuremath{F\!B}\xspace}
\def\etal  {{\it et~al.}}
\newcommand{\stat}{\ensuremath{\mathrm{(stat)}}\xspace}
\newcommand{\syst}{\ensuremath{\mathrm{(syst)}}\xspace}
\newcommand{\gev}{\ensuremath{\mathrm{\,Ge\kern -0.1em V}}\xspace}
\newcommand{\mev}{\ensuremath{\mathrm{\,Me\kern -0.1em V}}\xspace}
\newcommand{\gevc}{\ensuremath{{\mathrm{\,Ge\kern -0.1em V\!/}c}}\xspace}
\newcommand{\mevc}{\ensuremath{{\mathrm{\,Me\kern -0.1em V\!/}c}}\xspace}
\newcommand{\gevcc}{\ensuremath{{\mathrm{\,Ge\kern -0.1em V\!/}c^2}}\xspace}
\newcommand{\mevcc}{\ensuremath{{\mathrm{\,Me\kern -0.1em V\!/}c^2}}\xspace}

\graphicspath{{ps}}

\renewcommand{\arraystretch}{1.1}
\begin{document}

\preprint{\vbox{ \hbox{   }
}}

\title{\quad\\[0.5cm]
Search for {\boldmath $\CP$} violation in {\boldmath $D^0\to\pi^0\pi^0$} decays}

\noaffiliation
\affiliation{University of the Basque Country UPV/EHU, 48080 Bilbao}
\affiliation{University of Bonn, 53115 Bonn}
\affiliation{Budker Institute of Nuclear Physics SB RAS and Novosibirsk State University, Novosibirsk 630090}
\affiliation{Faculty of Mathematics and Physics, Charles University, 121 16 Prague}
\affiliation{University of Cincinnati, Cincinnati, Ohio 45221}
\affiliation{Deutsches Elektronen--Synchrotron, 22607 Hamburg}
\affiliation{Justus-Liebig-Universit\"at Gie\ss{}en, 35392 Gie\ss{}en}
\affiliation{Hanyang University, Seoul 133-791}
\affiliation{University of Hawaii, Honolulu, Hawaii 96822}
\affiliation{High Energy Accelerator Research Organization (KEK), Tsukuba 305-0801}
\affiliation{IKERBASQUE, Basque Foundation for Science, 48011 Bilbao}
\affiliation{Indian Institute of Technology Bhubaneswar, Satya Nagar 751007}
\affiliation{Indian Institute of Technology Guwahati, Assam 781039}
\affiliation{Indian Institute of Technology Madras, Chennai 600036}
\affiliation{Institute of High Energy Physics, Chinese Academy of Sciences, Beijing 100049}
\affiliation{Institute of High Energy Physics, Vienna 1050}
\affiliation{Institute for High Energy Physics, Protvino 142281}
\affiliation{INFN - Sezione di Torino, 10125 Torino}
\affiliation{Institute for Theoretical and Experimental Physics, Moscow 117218}
\affiliation{J. Stefan Institute, 1000 Ljubljana}
\affiliation{Kanagawa University, Yokohama 221-8686}
\affiliation{Institut f\"ur Experimentelle Kernphysik, Karlsruher Institut f\"ur Technologie, 76131 Karlsruhe}
\affiliation{Kavli Institute for the Physics and Mathematics of the Universe (WPI), University of Tokyo, Kashiwa 277-8583}
\affiliation{Korea Institute of Science and Technology Information, Daejeon 305-806}
\affiliation{Korea University, Seoul 136-713}
\affiliation{Kyungpook National University, Daegu 702-701}
\affiliation{\'Ecole Polytechnique F\'ed\'erale de Lausanne (EPFL), Lausanne 1015}
\affiliation{Faculty of Mathematics and Physics, University of Ljubljana, 1000 Ljubljana}
\affiliation{University of Maribor, 2000 Maribor}
\affiliation{Max-Planck-Institut f\"ur Physik, 80805 M\"unchen}
\affiliation{School of Physics, University of Melbourne, Victoria 3010}
\affiliation{Moscow Physical Engineering Institute, Moscow 115409}
\affiliation{Moscow Institute of Physics and Technology, Moscow Region 141700}
\affiliation{Graduate School of Science, Nagoya University, Nagoya 464-8602}
\affiliation{Kobayashi-Maskawa Institute, Nagoya University, Nagoya 464-8602}
\affiliation{Nara Women's University, Nara 630-8506}
\affiliation{Department of Physics, National Taiwan University, Taipei 10617}
\affiliation{H. Niewodniczanski Institute of Nuclear Physics, Krakow 31-342}
\affiliation{Nippon Dental University, Niigata 951-8580}
\affiliation{Niigata University, Niigata 950-2181}
\affiliation{University of Nova Gorica, 5000 Nova Gorica}
\affiliation{Osaka City University, Osaka 558-8585}
\affiliation{Pacific Northwest National Laboratory, Richland, Washington 99352}
\affiliation{Panjab University, Chandigarh 160014}
\affiliation{Peking University, Beijing 100871}
\affiliation{Punjab Agricultural University, Ludhiana 141004}
\affiliation{University of Science and Technology of China, Hefei 230026}
\affiliation{Soongsil University, Seoul 156-743}
\affiliation{Sungkyunkwan University, Suwon 440-746}
\affiliation{School of Physics, University of Sydney, NSW 2006}
\affiliation{Department of Physics, Faculty of Science, University of Tabuk, Tabuk 71451}
\affiliation{Tata Institute of Fundamental Research, Mumbai 400005}
\affiliation{Excellence Cluster Universe, Technische Universit\"at M\"unchen, 85748 Garching}
\affiliation{Toho University, Funabashi 274-8510}
\affiliation{Tohoku Gakuin University, Tagajo 985-8537}
\affiliation{Tohoku University, Sendai 980-8578}
\affiliation{Department of Physics, University of Tokyo, Tokyo 113-0033}
\affiliation{Tokyo Institute of Technology, Tokyo 152-8550}
\affiliation{Tokyo Metropolitan University, Tokyo 192-0397}
\affiliation{Tokyo University of Agriculture and Technology, Tokyo 184-8588}
\affiliation{CNP, Virginia Polytechnic Institute and State University, Blacksburg, Virginia 24061}
\affiliation{Wayne State University, Detroit, Michigan 48202}
\affiliation{Yamagata University, Yamagata 990-8560}
\affiliation{Yonsei University, Seoul 120-749}
\author{N.~K.~Nisar}\affiliation{Tata Institute of Fundamental Research, Mumbai 400005} 
\author{K.~Trabelsi}\affiliation{High Energy Accelerator Research Organization (KEK), Tsukuba 305-0801} 
\author{G.~B.~Mohanty}\affiliation{Tata Institute of Fundamental Research, Mumbai 400005} 
\author{T.~Aziz}\affiliation{Tata Institute of Fundamental Research, Mumbai 400005} 
\author{A.~Abdesselam}\affiliation{Department of Physics, Faculty of Science, University of Tabuk, Tabuk 71451} 
\author{I.~Adachi}\affiliation{High Energy Accelerator Research Organization (KEK), Tsukuba 305-0801} 
\author{H.~Aihara}\affiliation{Department of Physics, University of Tokyo, Tokyo 113-0033} 
\author{K.~Arinstein}\affiliation{Budker Institute of Nuclear Physics SB RAS and Novosibirsk State University, Novosibirsk 630090} 
\author{D.~M.~Asner}\affiliation{Pacific Northwest National Laboratory, Richland, Washington 99352} 
\author{V.~Aulchenko}\affiliation{Budker Institute of Nuclear Physics SB RAS and Novosibirsk State University, Novosibirsk 630090} 
% ...
\collaboration{The Belle Collaboration}

\begin{abstract}

We search for $\CP$ violation in neutral charm meson decays using
a data sample with an integrated luminosity of $966\invfb$ collected
with the Belle detector at the KEKB $e^+e^-$ asymmetric-energy collider. 
The asymmetry obtained in the rate of $D^0$ and $\Dzb$ decays to the 
$\pi^0\pi^0$ final state, $[-0.03\pm0.64\stat\pm0.10\syst]\%$, is consistent 
with no $\CP$ violation. This constitutes an
order of magnitude improvement over the existing result. We also
present an updated measurement of the $\CP$ asymmetry in the $D^0\to\KS
\pi^0$ decay: $A_{\CP}(D^0\to\KS\pi^0) = [-0.21\pm0.16\stat\pm0.07\syst]\%$.

\end{abstract}

\maketitle

{\renewcommand{\thefootnote}{\fnsymbol{footnote}}}
\setcounter{footnote}{0}

Within the Standard Model (SM), $\CP$ violation in charm
decays~\cite{Bigi:2011re,Isidori:2011qw,Brod:2011re} is expected
to be very small and thus challenging to observe experimentally.
Observing such $\CP$ violation could indicate new physics. The
$D^0\to\pi^0\pi^0$ decay proceeds via a singly Cabibbo-suppressed
(SCS) amplitude, which is expected to have enhanced interference
with new physics amplitudes. Such interference could generate a large
$\CP$ violation effect. An early observation by
LHCb~\cite{Aaij:2011in} suggested a $3.5$ standard deviation
($\sigma$) effect on the difference of direct $\CP$ asymmetries
($\Delta A_{\CP}$) between $D^0\to K^+K^-$ and $D^0\to\pi^+\pi^-$
decays that was later supported by the CDF
experiment~\cite{Collaboration:2012qw}. At the end of 2012, the
world average~\cite{hfag} for $\Delta A_{\CP}$ was $4.6\,\sigma$
away from zero. This triggered much theoretical 
activity~\cite{Lenz:2013pwa} in an attempt to explain the effect.

In the SM, $\CP$ violation in SCS charm decays arises
due to interference between the tree and loop (penguin) amplitudes and
is suppressed by $\order(V_{cb}V_{ub}/V_{cs}V_{us})\sim 10^{-3}$, where
$V_{ij}$ are the elements of the Cabibbo-Kobayashi-Maskawa (CKM)
matrix~\cite{ckm}. The uncertainties on these order-of-magnitude estimates 
are, however, large~\cite{Brod:2011re}. 
Although a large $\Delta A_{\CP}$ could be explained by non-SM physics, it 
may be simply due to an unexpectedly enhanced $\CP$-violating SM $c \to u$ 
penguin amplitude. In the latter case, one expects fractional-percent
$\CP$ asymmetries in other SCS two-body decays such as $D^0\to\pi^0
\pi^0$~\cite{Cheng:2012wr,Bhattacharya:2012ah,Grossman:2012eb,Hiller:2012xm}.
Recently, new measurements of $\Delta A_{\CP}$ have been
performed~\cite{LHCb-CONF-2013-003,Aaij:2013bra}, and the current world
average is $2.3\,\sigma$ away from zero~\cite{hfag}. The only search
for $\CP$ violation in $D^0\to\pi^0\pi^0$ was performed by the CLEO
Collaboration using $13.7\invfb$ of data~\cite{Bonvicini:2000qm};
the result was $ A_{\CP}=(+0.1\pm4.8)\%$.

In this Letter, we measure the time-integrated $\CP$-violating asymmetry
($A_{\CP}$) in neutral charm meson decays to a pair of neutral pions, 
$D^0\to\pi^0\pi^0$~\cite{conjugate}. We also update our $D^0\to\KS\pi^0$ 
result~\cite{Ko:2011ey} using Belle's full data sample. The SM predicts 
a nonzero $\CP$ asymmetry in final states containing a neutral kaon due 
to $K^0$-$\Kzb$ mixing, even if no $\CP$ violating phase exists in the charm
decay amplitudes. The expected magnitude for this type of asymmetry 
is $A_{\CP}^{\Kzb}=(-0.339\pm0.007)\%$~\cite{Ko:2012pe}.

The charge of the accompanying low-momentum or ``slow'' pion, 
$\pi^+_{s}$, in the decay 
$\Dstarp\to D^0\pi^+_{s}$~\cite{conjugate} identifies the flavor of the neutral
charm meson (whether it is a $D^0$ or a $\Dzb$) at its production. The
measured asymmetry 
\begin{equation}
A_{\rm rec}=\frac{N_{\rm rec}^{\Dstarp\to D^0\pi^+_{s}}-N_{\rm rec}^{\Dstarm\to\Dzb\pi^-_{s}}}
{N_{\rm rec}^{\Dstarp\to D^0\pi^+_{s}}+N_{\rm rec}^{\Dstarm\to\Dzb\pi^-_{s}}},
\label{eq_asym_rec}
\end{equation} 
where $N_{\rm rec}$ is the number of reconstructed signal events, includes 
three contributions: the underlying $\CP$ asymmetry $A_{\CP}$,
the forward-backward asymmetry ($A_{\FB}$) due to $\gamma$-$Z^0$ interference 
in $e^+e^-\to\ccbar$ and higher order QED effects~\cite{qed_ref}, and the
detection asymmetry between positively and negatively charged pions
($A^{\pi_s}_{\epsilon}$). 
The last contribution depends on the transverse momentum $p_{T}^{\pi_s}$
and polar angle $\theta^{\pi_s}$ of the slow pion and is independent of
the $D^0$ decay final state. To estimate $A^{\pi_s}_{\epsilon}$, we use
the Cabibbo-favored decay $D^0\to K^-\pi^+$ (``untagged'') and
$\Dstarp\to D^0\pi^+_s\to K^-\pi^+\pi^+_s$ (``tagged''), and we assume
the same $A_{\FB}$ for $\Dstarp$ and $D^0$ mesons~\cite{Staric:2008rx}. By
subtracting the measured asymmetries in these two decay modes, we directly
obtain the $A^{\pi_s}_{\epsilon}$ correction factor [$\order(0.1\%)$]. After 
$A_{\rm rec}$ is corrected for $A^{\pi_s}_{\epsilon}$, one is left with 
\begin{equation}
A_{\rm rec}^{\rm cor} = A_{\CP} + A_{\FB} (\cos\theta^*).
\label{eq_rec_corr}
\end{equation} 
While $A_{\CP}$ is independent of all kinematic
variables, $A_{\FB}$ is an odd function of the cosine of the $\Dstarp$ polar
angle, $\theta^*$, in the center of mass (c.m) system. 
We thus extract $A_{\CP}$ and $A_{\FB}$ using
\begin{equation}
A_{\CP}=[A_{\rm rec}^{\rm cor}(\cos\theta^*) + A_{\rm rec}^{\rm cor}(-\cos\theta^*)]/2,
\label{eq_acp}
\end{equation}
and
\begin{equation}
A_{\FB}=[A_{\rm rec}^{\rm cor}(\cos\theta^*) - A_{\rm rec}^{\rm cor}(-\cos\theta^*)]/2.
\label{eq_afb}
\end{equation} 

The analysis is based on a data sample corresponding to an integrated
luminosity of $966\invfb$ collected at the $\Upsilon(nS)$ resonances 
($n = 1, 2, 3, 4, 5$) or $60\mev$ below the $\Y4S$ resonance with 
the Belle detector~\cite{belle} at the KEKB asymmetric-energy $e^+e^-$ 
collider~\cite{kekb}. In the following, the samples taken at or below 
the $\Y4S$ resonance will be referred to as $\Y4S$, while the sample 
recorded at the $\Y5S$ is considered separately. The detector components
relevant for our study are: a tracking system comprising a silicon
vertex detector (SVD) and a 50-layer central drift chamber (CDC), 
a particle identification (PID) system that consists of a barrel-like
arrangement of time-of-flight scintillation counters (TOF) and 
an array of aerogel threshold Cherenkov counters (ACC), and a CsI(Tl)
crystal-based electromagnetic calorimeter (ECL). All these components
are located inside a superconducting solenoid coil that provides a
$1.5$\,T magnetic field.

We use Monte Carlo (MC) simulated events representing a luminosity six times 
that of the data to devise selection criteria and investigate possible
sources of background. The selection optimization is performed by minimizing
the expected statistical error on $A_{\rm rec}$, where the branching fraction
of $D^0\to\pi^0\pi^0$ is set to $8\times10^{-4}$~\cite{Lees:2011qz} in MC 
simulations. 
The level of background is obtained by appropriately scaling the number 
of events observed in a data sideband of the reconstructed $\Dstar$ mass.

Candidates for the $\KS\to\pi^+\pi^-$ decay are formed from pairs of
oppositely charged tracks having a reconstructed invariant mass within
$9\mevcc$ (about three times the experimental resolution) 
of the nominal $\KS$ mass~\cite{PDG}. The $\KS$ candidates are also 
required to satisfy the criteria described in Ref.~\cite{Chen:2005dra}
to ensure that their decay vertices are displaced from the interaction 
point (IP).
We reconstruct neutral pion candidates from pairs of electromagnetic showers
in the ECL that are not matched to any charged track. Showers in the barrel
(end-cap) region of the ECL must exceed $60$ $(100)\mev$ to be considered
as a $\pi^0$ daughter candidate. The invariant mass of the $\pi^0$ candidate
must lie within $25\mevcc$ (about four times the experimental 
resolution) of the known $\pi^0$ mass~\cite{PDG}. The $\pi^0$
momentum is required to be greater than $640$ $(540)\mevc$ for the data
sample taken at the $\Y4S$ ($\Y5S$) resonance.

Reconstructed $\pi^0$ and $\KS$ candidates are kinematically constrained 
to the nominal $\pi^0$ and $\KS$ mass values and combined to form 
$D^0\to\KS\pi^0$ and $D^0\to\pi^0\pi^0$ candidates. 
For the former, we retain the $D^0$ candidates having an
invariant mass in the range $1.750<M<1.950\gevcc$, whereas for the
latter the range is $1.758<M<1.930\gevcc$ in order to suppress
background from $D^0\to\KS(\pi^0\pi^0)\pi^0$. 

We require $\pi^+_s$ candidates to originate from near the IP by restricting 
their impact parameters along and perpendicular to the $z$ axis to be less 
than $3\cm$ and $1\cm$, respectively. The $z$ axis is defined to be the 
direction opposite the $e^+$ beam. We do not impose any requirement on 
the number of SVD hits but require that the ratio of PID likelihoods, 
${\cal L}_\pi/({\cal L}_\pi+{\cal L}_K)$, be greater than~$0.4$. 
Here, ${\cal L}_\pi$ $({\cal L}_K)$ is the likelihood of a track being a
pion (kaon) and is calculated using specific ionization from the CDC,
time-of-flight information from the TOF and the number of photoelectrons in
the ACC. With the above PID requirement, the pion identification efficiency
is above $95\%$ with a kaon misidentification probability below $5\%$. 

$\Dstarp$ candidates are reconstructed by combining the $\pi^+_s$ with a $D^0$
candidate and requiring that the resultant $\Delta M$ value lies in the range
$[0.14,$ $0.16]\gevcc$, where $\Delta M\equiv M(\Dstarp)-M(D^0)$. In order to
improve the $\Delta M$ resolution, the $\pi^+_s$ is constrained to originate
from the IP. The sideband used for the selection optimization is 
$0.15<\Delta M<0.16\gevcc$. $D$ mesons produced in $B$ meson decays are 
rejected by requiring that the $\Dstarp$ candidates have a CM momentum 
greater than $2.5\gevc$ and $3.1\gevc$, respectively, for data taken near 
the $\Y4S$ and $\Y5S$ resonance. This requirement also significantly
reduces combinatorial background.

After applying all selection criteria, we find that about $6\%$ of the total 
$\Dstar\to D(\pi^0\pi^0)\pi_s$ events contain multiple candidates, of which 
about half are due to a misreconstructed $\pi^0$ and about half due to  
a misreconstructed $\pi_s$. We select a single $D^0$ candidate per event 
by choosing that which has the smallest $\chi^2_{\rm BCS}$. This quantity 
is defined as:
\begin{equation}
\chi^2_{\rm BCS}=\sum\chi^2_{\pi^0}+\left[\frac{M(D^0)-m_{D^0}}{\sigma_M}\right]^2,
\end{equation}
where $\chi^2_{\pi^0}$ is the $\pi^0$ mass-constrained fit statistic, 
$\sigma_M$ is the uncertainty on the reconstructed $D$ mass as determined 
from MC simulations, and $m_{D^0}$ is the nominal $D^0$ mass~\cite{PDG}. 
In case the $D^0$ candidate is common to more than one $\Dstar$ candidate, 
we select the one having the slow pion with the smallest impact parameter 
perpendicular to the $z$ axis. According to MC simulation, this procedure 
identifies the correct $\Dstar$ candidate among multiple candidates
about 74\% of the time.

\begin{figure}[htbp]
\begin{center}
  \begin{tabular}{cc}  
        \includegraphics[width=0.24\textwidth]{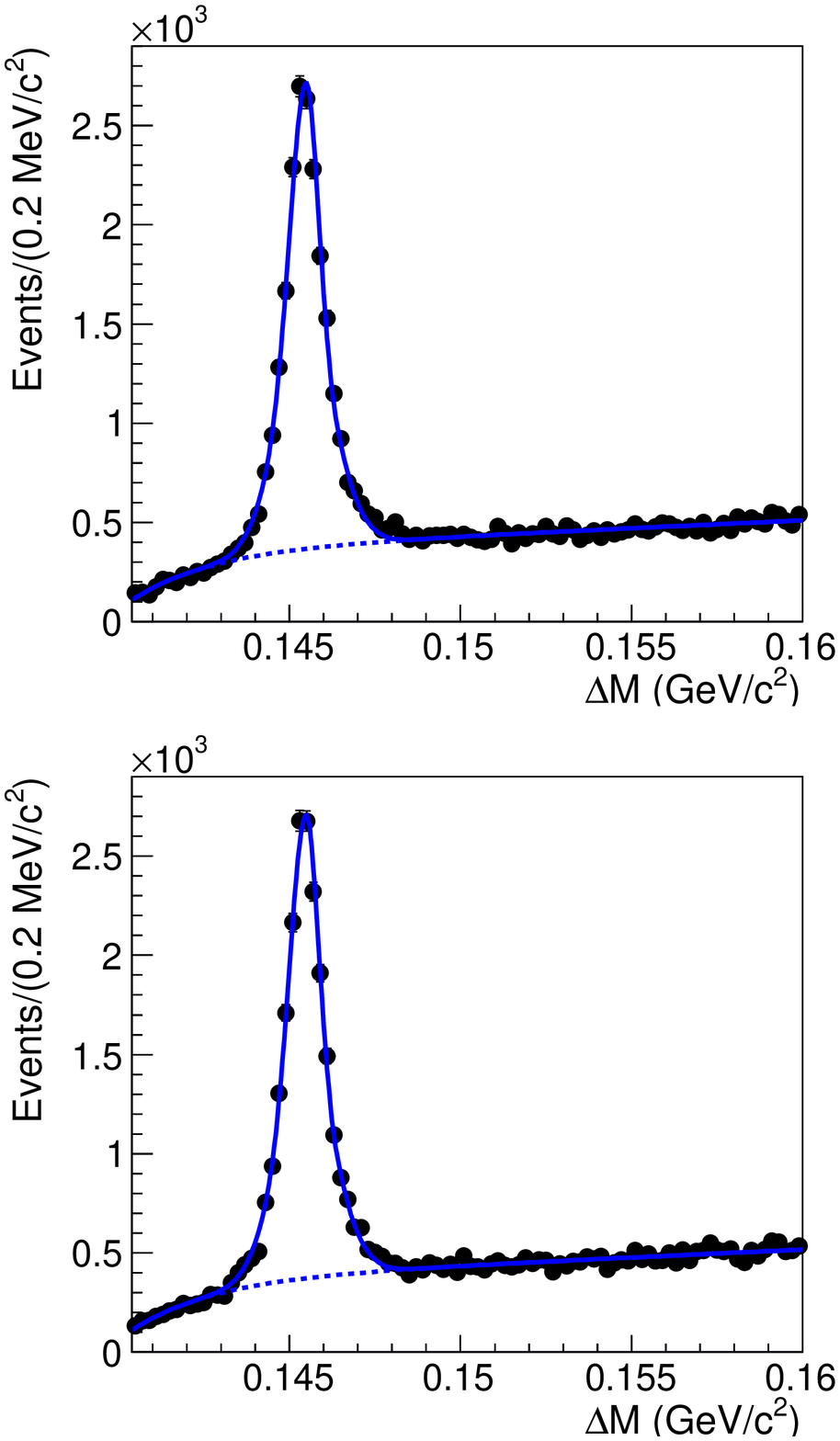} 
        \includegraphics[width=0.24\textwidth]{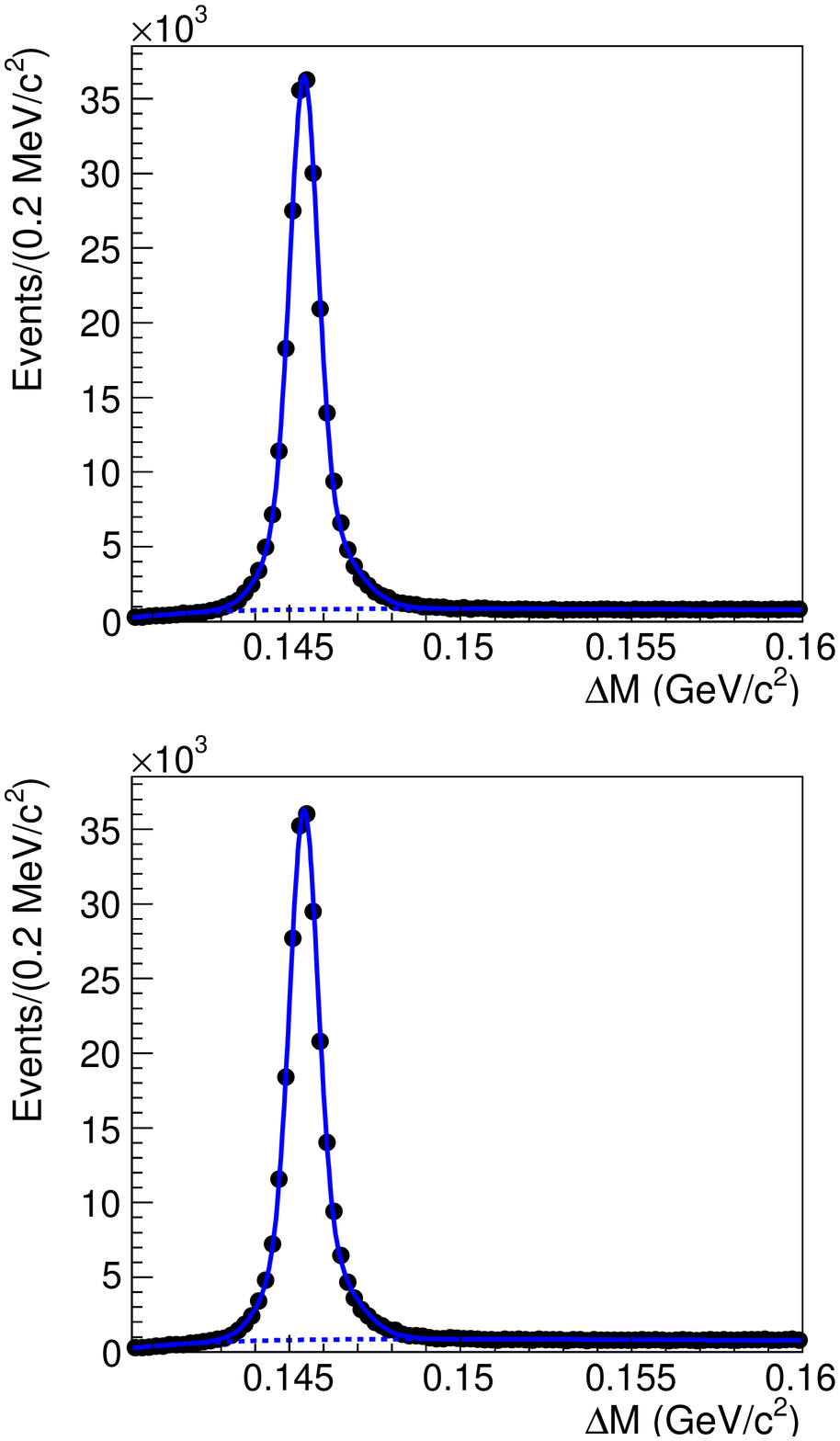} 
\end{tabular}
\caption{Distributions of the mass difference $\Delta M$ for the 
$\pi^0\pi^0$ (left) and $\KS\pi^0$ (right) final states. Top (bottom)
plots are for the $\Dstarp$ ($\Dstarm$) sample. Points with error bars
are the data, the solid curves show the results of the fit and the
dashed curves are the background predictions.}
\label{fig:dm_kspi_pipi}
\end{center}
\end{figure}
Figure~\ref{fig:dm_kspi_pipi} shows the $\Delta M$ distributions of
event candidates in the two decay modes. We describe the signal shapes 
by the sum of symmetric and asymmetric Gaussian functions with a 
common mean. The background shapes are modeled with a threshold function 
as $(x-m_{\pi})^\alpha\exp[-\beta(x-m_{\pi})]$, where $m_{\pi}$ is the 
nominal charged pion mass~\cite{PDG}, and $\alpha$ and $\beta$ are  
shape parameters. The asymmetry $A_{\rm rec}$ and the sum of the $\Dstarp$ 
and $\Dstarm$ yields are obtained from a simultaneous fit to their $\Delta M$ 
distributions. The parameters common in the fit are: (for signal) the common 
mean, the widths of the symmetric and the asymmetric Gaussian functions, 
and the relative fraction of the two functions, 
(for background) $\alpha$ and $\beta$. 
The signal yield for $D^0\to\pi^0\pi^0$ is $34\,460\pm273$ events and 
$A_{\rm rec}=(+0.29\pm0.64)\%$. For the $D^0\to\KS\pi^0$ case, the signal 
yield is $466\,814\pm773$ events and $A_{\rm rec}=(+0.29\pm0.15)\%$. 

\begin{figure}[htbp]
\begin{center}
\begin{tabular}{cc}  
 \includegraphics[width=0.24\textwidth]{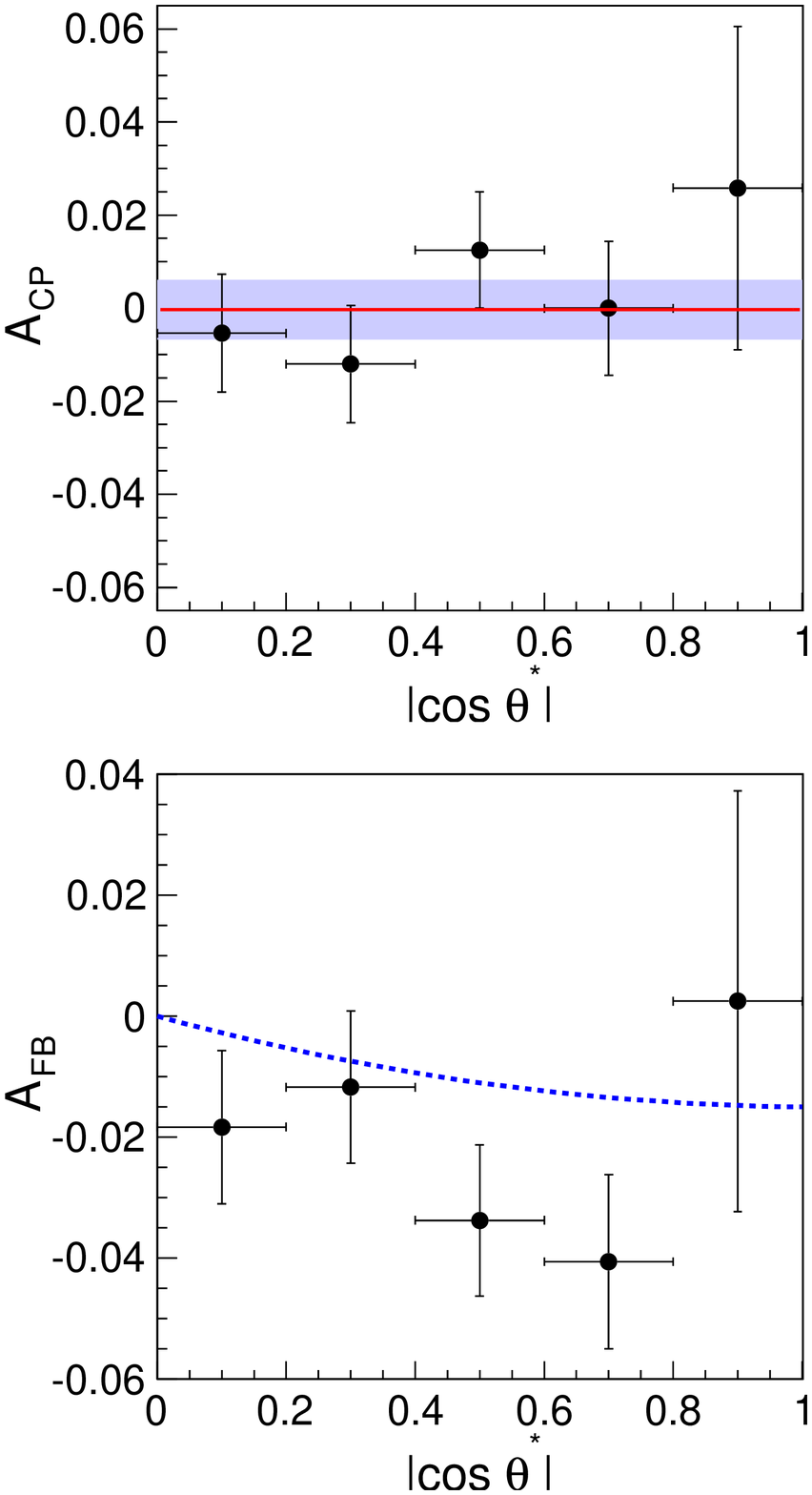}
 \includegraphics[width=0.24\textwidth]{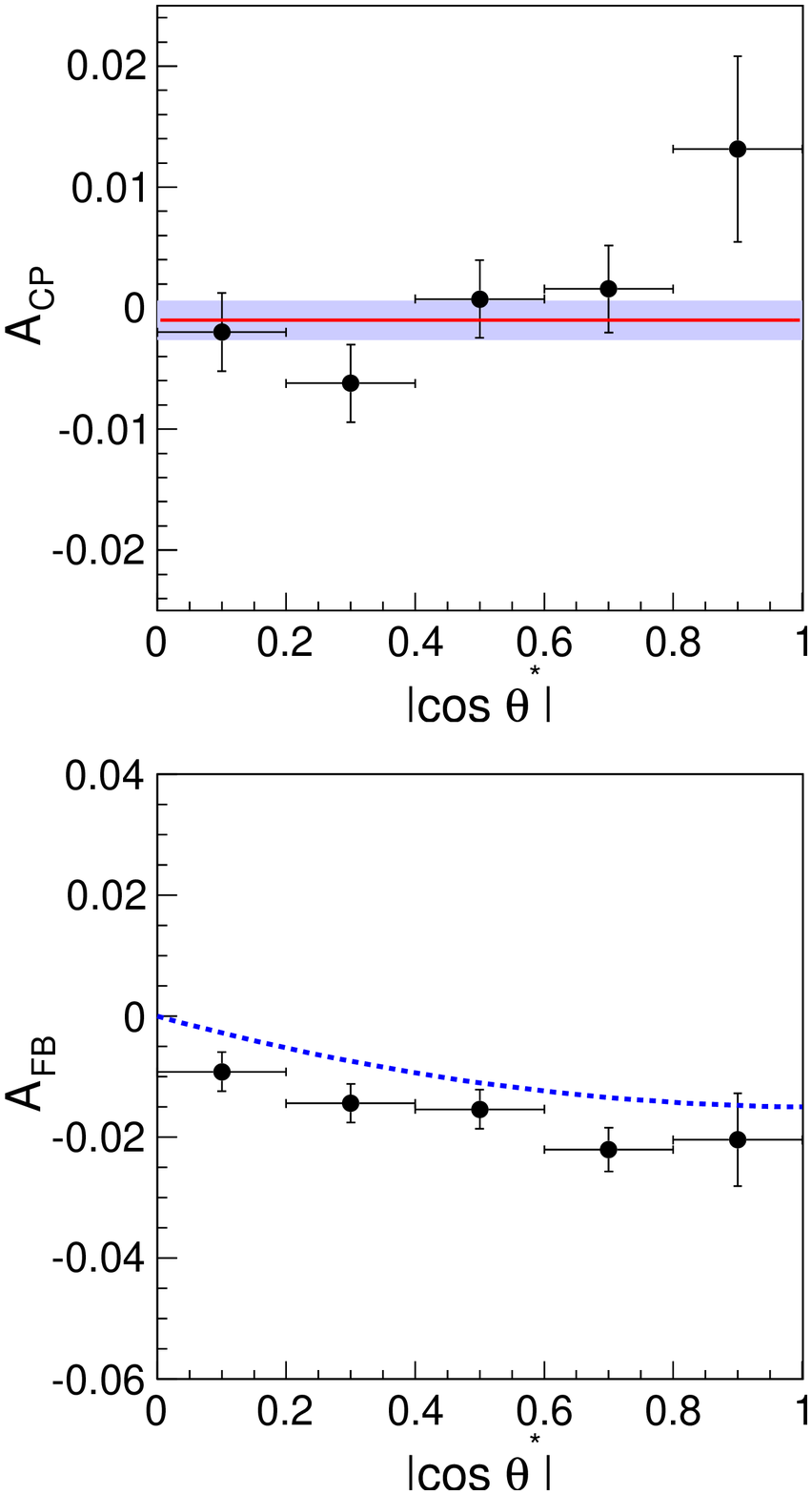}
\end{tabular}
\caption{(color online) $\CP$ violation asymmetry $A_{\CP}$ (top) and 
forward-backward asymmetry $A_{\FB}$ (bottom) values as a function of 
$|\cos\theta^*|$.
Plots on the left (right) are for the $\pi^0\pi^0$ ($\KS\pi^0$) final
state. The solid red lines represent the central values obtained from a
least-square minimization, the blue regions for the $A_{\CP}$ plots
show the $1\sigma$ interval, and the dashed blue curves for the $A_{\FB}$
plots show the leading-order prediction for $A_{\FB}(e^+e^-\to\ccbar)$.}
\label{fig:acp_kspi_pipi}
\end{center}
\end{figure}

The data samples shown in Fig.~\ref{fig:dm_kspi_pipi} are divided
into ten bins of $\cos\theta^*$, seven bins of $p_{T}^{\pi_s}$, and 
eight bins of $\cos\theta^{\pi_s}$. For each 3D bin, a simultaneous fit
analogous to the one used for the full sample is performed, and 
the asymmetry obtained for each bin is 
corrected by the corresponding $A^{\pi_s}_{\epsilon}$ obtained in 
Ref.~\cite{Ko:2011ey}.
Due to limited statistics, the shape for the $D^0\to\pi^0\pi^0$ signal 
in a bin of [$p_{T}^{\pi_s}$, $\cos\theta^{\pi_s}$] is taken from the
larger $\KS\pi^0$ sample. We account for small differences between
the two samples using MC simulations. 
Bins with fewer than $30$ events, which correspond to only $2\%$ of
the total statistics in the $D^0\to\pi^0\pi^0$ sample, are removed 
from the $A_{\CP}$ estimation. A weighted average over the 
[$p_{T}^{\pi_s}$, $\cos\theta^{\pi_s}$] bins having the same 
$\cos\theta^*$ value is then performed, and $A_{\CP}$ and $A_{\FB}$ are extracted
from Eqs.~(\ref{eq_acp}) and (\ref{eq_afb}), respectively. This procedure has 
been verified with six sets of generic MC samples, each of similar size 
as the data; the resulting $A_{\CP}$ values were found to be in agreement with 
the generated values. Figure~\ref{fig:acp_kspi_pipi} shows $A_{\CP}$ and
$A_{\FB}$ as a function of $|\cos\theta^*|$ obtained for the two data samples.
From the weighted average over the $|\cos\theta^*|$ bins, we obtain
\begin{eqnarray}
A_{\CP}(\pi^0\pi^0)=(-0.03\pm0.64)\%, \\
A_{\CP}(\KS\pi^0)=(-0.10\pm0.16)\%, 
\end{eqnarray}
where the uncertainties are statistical only, with a reduced $\chi^2$ of 1.7 
and 0.7, respectively. The observed $A_{\FB}$ values decrease with 
$|\cos\theta^*|$ as expected but are somewhat lower than the leading order 
QED prediction~\cite{qed_ref}. Higher-order corrections are expected to lower
the theoretical prediction, which would bring it into better agreement with
our data. 

We identify three significant sources of systematic uncertainty (see 
Table~\ref{tab:syst}). The first is due to the 
uncertainty in the signal shapes, which, in the case of $D^0\to\pi^0\pi^0$,  
is dominated by the statistics of the calibration mode $D^0\to\KS\pi^0$. 
The second is the slow pion efficiency correction. We estimate its
contribution by varying $A^{\pi_s}_{\epsilon}$ by its statistical error
in each of the $7\times 8$ bins of [$p_{T}^{\pi_s}$, $\cos\theta^{\pi_s}$].
The third is the $A_{\CP}$ extraction 
procedure itself and is obtained by varying the binning in $|\cos\theta^*|$.
For the $D^0\to\KS\pi^0$ channel, we correct for a non-vanishing asymmetry 
originating from the different strong interaction of $K^0$ and $\Kzb$ mesons 
with nucleons of the detector material, estimated to be $-0.11\%$ in 
Ref.~\cite{Ko:2010mk}, and assign an additional systematic uncertainty
of $0.01\%$. 
Finally, we add these individual contributions in quadrature to obtain 
the total systematic uncertainty. The result is $0.10\%$ $(0.07\%)$ for 
the $\pi^0\pi^0$ ($\KS\pi^0$) sample.
\begin{table}[tbh]  
\caption{Summary of systematic uncertainties (\%) in $A_{\CP}$. }
\label{tab:syst}
\renewcommand{\arraystretch}{1.1}   
\begin{ruledtabular}   
\begin{tabular}{l|cc}    
Source & $\pi^0\pi^0$ & $\KS\pi^0$ \\   
\hline                    
Signal shape               & $\pm 0.03$ & $\pm 0.01$  \\  
Slow pion correction       & $\pm 0.07$ & $\pm 0.07$  \\  
$A_{\CP}$ extraction method & $\pm 0.07$ & $\pm 0.02$ \\
$K^0\!/\Kzb$-material effects & --  & $\pm 0.01$ \\
\hline
Total                      & $\pm 0.10$ & $\pm 0.07$  \\      
\end{tabular}
\end{ruledtabular}
\end{table}

In summary, we have measured the time-integrated $\CP$-violating asymmetry 
$A_{\CP}$ in the $D^0\to\pi^0\pi^0$ decay using $966\invfb$ of data. 
After correcting for the detector-induced asymmetries with a precision 
of $0.07\%$ by using the tagged and untagged $D^0\to K^-\pi^+$ decays, 
we obtain:
\begin{eqnarray}
A_{\CP}(D^0\to\pi^0\pi^0)=(-0.03\pm0.64\pm0.10)\%, 
\end{eqnarray}
where the uncertainties are statistical and systematic, respectively.
The measured $\CP$ asymmetry has an order of magnitude better precision 
than the previous result~\cite{Bonvicini:2000qm} and shows no evidence 
for $\CP$ violation. We also measure:
\begin{eqnarray}
A_{\CP}(D^0\to\KS\pi^0)=(-0.21\pm0.16\pm0.07)\%, 
\end{eqnarray}
which supersedes our earlier result~\cite{Ko:2011ey}. After subtracting
$\CP$ violation due to $K^0$-$\Kzb$ mixing,
$(-0.339\pm0.007)\%$\cite{Ko:2012pe}, the $\CP$ asymmetry in
$D^0\to\Kzb\pi^0$ decay is found to be $(+0.12\pm0.16\pm0.07)\%$
that is consistent with no $\CP$ violation.
\\

We thank the KEKB group for excellent operation of the
accelerator; the KEK cryogenics group for efficient solenoid
operations; and the KEK computer group, the NII, and 
PNNL/EMSL for valuable computing and SINET4 network support.  
We acknowledge support from MEXT, JSPS and Nagoya's TLPRC (Japan);
ARC and DIISR (Australia); FWF (Austria); NSFC (China); MSMT (Czechia);
CZF, DFG, and VS (Germany); DST (India); INFN (Italy); 
MOE, MSIP, NRF, GSDC of KISTI, BK21Plus, and WCU (Korea);
MNiSW and NCN (Poland); MES and RFAAE (Russia); ARRS (Slovenia);
IKERBASQUE and UPV/EHU (Spain); 
SNSF (Switzerland); NSC and MOE (Taiwan); and DOE and NSF (USA).


\begin{thebibliography}{99}

\bibitem{Bigi:2011re}
 I.I. Bigi, A. Paul, and S. Recksiegel, 
 J. High Energy Phys. 06 (2011) 089.

\bibitem{Isidori:2011qw}
 G. Isidori, J.F. Kamenik, Z. Ligeti, and G. Perez,
 Phys.\ Lett.\ B {\bf 711}, 46 (2012).

\bibitem{Brod:2011re}
 J. Brod, A. Kagan, and J. Zupan, Phys.\ Rev.\ D {\bf 86}, 014023 (2012).

\bibitem{Aaij:2011in}
 R. Aaij \etal\ (LHCb Collaboration), 
 Phys.\ Rev.\ Lett.\ {\bf 108}, 111602 (2012).

\bibitem{Collaboration:2012qw}
 T. Aaltonen \etal\ (CDF Collaboration), 
 Phys.\ Rev.\ Lett.\ {\bf 109}, 111801 (2012).

\bibitem{hfag}
 Y. Amhis \etal\ (Heavy Flavor Averaging Group), arXiv:1207.1158 
 and online update at {\tt http://www.slac.stanford.edu/xorg/hfag}.

\bibitem{Lenz:2013pwa}
 A. Lenz, arXiv:1311.6447[hep-ph] and references within.

\bibitem{ckm}
 N. Cabibbo, Phys.\ Rev.\ Lett.\ {\bf 10}, 531 (1963);
 M. Kobayashi and T. Maskawa, Prog.\ Theor.\ Phys.\ {\bf 49}, 652 (1973).

\bibitem{Cheng:2012wr}
 H-Y. Cheng and C-W. Chiang, Phys.\ Rev.\ D {\bf 85}, 034036 (2012);
{\bf 85}, 079903(E) (2012);
{\bf 86}, 014014 (2012).

\bibitem{Bhattacharya:2012ah}
 B. Bhattacharya, M. Gronau, and J.L. Rosner, Phys.\ Rev.\ D {\bf 85}, 054014 (2012).

\bibitem{Grossman:2012eb}
 Y. Grossman, A. Kagan, and J. Zupan, Phys.\ Rev.\ D {\bf 85}, 114036 (2012).

\bibitem{Hiller:2012xm}
 G. Hiller, M. Jung, and S. Schacht, Phys.\ Rev.\ D {\bf 87}, 014024 (2013).

\bibitem{LHCb-CONF-2013-003}
(LHCb Collaboration), LHCb-CONF-2013-003.

\bibitem{Aaij:2013bra}
 R. Aaij \etal\ (LHCb Collaboration), 
 Phys.\ Lett.\ B {\bf 723}, 33 (2013).

\bibitem{Bonvicini:2000qm}
 G. Bonvicini \etal\ (CLEO Collaboration), 
 Phys.\ Rev.\ D {\bf 63}, 071101 (2001).

\bibitem{conjugate}
Throughout this Letter, the charge-conjugate decay mode is implied unless stated otherwise.

\bibitem{Ko:2011ey}
 B.R. Ko \etal\ (Belle Collaboration), 
 Phys.\ Rev.\ Lett.\ {\bf 106}, 211801 (2011).

\bibitem{Ko:2012pe}
 B.R. Ko \etal\ (Belle Collaboration), 
 Phys.\ Rev.\ Lett.\ {\bf 109}, 021601 (2012);  
 {\bf 109}, 119903(E) (2012)].
\bibitem{qed_ref}
 F.A. Berends, K.J.F. Gaemers, and R. Gastmans, Nucl.\ Phys.\ B {\bf 63}, 381 (1973); 
 R.W. Brown, K.O. Mikaelian, V.K. Cung, and E.A. Paschos, Phys.\ Lett.\ {\bf 43B}, 403 (1973);
 R.J. Cashmore, C.M. Hawkes, B.W. Lynn, and R.G. Stuart, Z Phys.\ C {\bf 30}, 125 (1986).

\bibitem{Staric:2008rx}
 M. Staric \etal\ (Belle Collaboration), 
 Phys.\ Lett.\ B {\bf 670}, 190 (2008).

\bibitem{belle}
 A. Abashian \etal\ (Belle Collaboration), Nucl.\ Instrum.\ Methods 
 Phys.\ Res., Sect.\ A {\bf 479}, 117 (2002); also, see the detector section
 in J. Brodzicka \etal, Prog.\ Theor.\ Exp.\ Phys., 04D001 (2012).

\bibitem{kekb} 
 S. Kurokawa and E. Kikutani,
 Nucl.\ Instrum.\ Methods Phys.\ Res., Sect.\ A {\bf 499}, 1 (2003),
 and other papers in this volume;
 T. Abe \etal, Prog.\ Theor.\ Exp.\ Phys., 03A001 (2013) and following
 articles up to 03A011.

\bibitem{Lees:2011qz}
 J.P. Lees \etal\ (\babar\ Collaboration), 
 Phys.\ Rev.\ D {\bf 85}, 091107 (2012).

\bibitem{PDG}
 J. Beringer \etal\ (Particle Data Group),
 Phys.\ Rev.\ D {\bf 86}, 010001 (2012).

\bibitem{Chen:2005dra}
 K.-F. Chen \etal\ (Belle Collaboration), 
 Phys.\ Rev.\ D {\bf 72}, 012004 (2005).

\bibitem{Ko:2010mk}
 B.R. Ko, E. Won, B. Golob, and P. Pakhlov, 
 Phys.\ Rev.\ D {\bf 84}, 111501 (2011).

\end{thebibliography}
\end{document}